\documentclass[12pt]{iopart}

\begin{document}

\title[]{The GUP effect on tunneling of massive vector bosons from the 2+1 dimensional black hole}

\author{Ganim Gecim and Yusuf Sucu}

\address{Department of Physics, Faculty of Science, Akdeniz University, 07058 Antalya, Turkey}
\ead{gecimganim@gmail.com and ysucu@akdeniz.edu.tr}

\vspace{10pt}

\begin{abstract}
In this study, the Generalized Uncertainty Principle (GUP) effect on the Hawking radiation formed by tunneling of a massive vector boson particle from the
$2+1$ dimensional New-type Black Hole was investigated. We used modified massive vector boson equation based on the GUP. Then, the Hamilton-Jacobi quantum tunneling approach was used to work out the tunneling probability of the massive vector boson particle and, Hawking temperature of the black hole. Due to the GUP effect, the modified Hawking temperature was found to depend on the black hole properties, the AdS$_{3}$ radius, and on the energy, mass and total angular momentum of the tunneling massive vector boson. In the light of these results, we also observed that modified Hawking temperature increases by the total angular momentum the particle while it decreases by the energy and mass of the particle, and the graviton mass. Also, in the context of the GUP, we see that the Hawking temperature due to the tunnelling massive vector boson is completely different from both that of the spin-0 scalar and the spin-1/2 Dirac particles obtained in the previous study. We also calculate the heat capacity of the black hole using the modified Hawking temperature and then discuss influence of the GUP on the stability of the black hole.
\end{abstract}

%
%
%
%
%

\section{Introduction} \label{intro}

It was theoretically proved by Hawking that a black hole can emit thermal radiation as a result of quantum tunneling process of the particles created by the quantum fluctuation near the event horizon \cite{1c,1d,1e,1bc}. This suggestion has an important influence in understanding the quantization of black hole and gravity \cite{Rob,Gid}. Since the Hawking's discovery, many different methods were developed to calculate black hole radiation, known as Hawking radiation, in the literature. For instance, the Hamilton-Jacobi approach is an effective way to study the Hawking radiation as a quantum tunneling process of particles from black holes. In this context, Hawking radiation is seen to be extensively investigated in the literature in the context of quantum tunneling of a point-like elementary particle \cite{2,3,4,5,6,7,8,9,10,11,12,13,13a,14,14a,18,18a,18b,18c,19,19a,19c,19d,19e,14cd,GY1,GY2,GY3}.
These studies show that Hawking radiation of a black hole must be completely independent from the mass, total angular momentum, energy, charge of a tunneling point particle.

On the other hand, alternative approaches about quantum gravity predict the presence of a
minimal observable length in Planck scale \cite{21,22,23,24,kem,ali,taw1,taw2}. The existence of such length leads
to the GUP, a generalization of the standard Heisenberg uncertainty principle \cite{22,magg,scar,ghan}. Effects of the GUP on the standard quantum mechanical problems \cite{Brau,Pedram1}, on the early and late times of the universe \cite{nozari1,vakili,Ali0,majumder}, and on thermodynamic properties of black holes \cite{sad,myung,ali2} have been extensively discussed in the literature. To calculate the modified Hawking radiation in the GUP framework by Hamilton-Jacobi approach, Klein-Gordon equation for spin-0 particle, Dirac equation for spin-1/2 particle, and massive $W^{\mp}$-boson equation for spin-1 particle are modified \cite{jan,24aa,GWS}. It shows that the modified Hawking temperature depends not only on properties of black hole,  but also on the properties of  tunneling particle, such as mass, angular (orbital+spin) momentum, energy, charge
\cite{sagh,24c,24d,24f,24g,24h,24j,24k,24l,24m,24n,liz,sakalli,vagenas}. Furthermore, it has been proved that
the spin-0 and spin-1/2 particles tunnel through the horizon of the $2+1$ dimensional black holes in the same way
in the absence of the GUP effect. On the other hand, they tunnel differently in the presence of the GUP effect \cite{GY4,GY5}. Therefore, our plan was to figure out the effect of the GUP on the Hawking temperature of the $2+1$ dimensional new-type black hole by using process of tunneling of the massive spin-1 particle described by the relativistic equation derived from the quantization of classical zitterbewegung model \cite{16a,16b}. This will provide a comparison of the results of the tunneling process of the different particles types from the new-type black hole under conditions with and without the GUP effect \cite{GY1,GY2,GY3,GY5}. The stability of the black hole will also be discussed in the context of the GUP.

The paper is organized as follows: in the following section, we modify the massive vector boson equation in the framework of
the GUP. After that, in section 3, we introduce the new-type black hole, and then calculate its modified Hawking temperature by using the Hamilton-Jacobi approach. Moreover, using the modified Hawking temperature, the heat capacity of the black hole is calculated for analyzing the stability of the black hole. In section 4, we summarize the results.

\section{Modified Vector Boson Equation}\label{Vboson}

In a curved spacetime background, the massive vector boson equation can be written as in \cite{16a}
\begin{equation}
i\beta^{\mu}(x)\left[\partial_{\mu}-\Sigma_{\mu}(x)\right]\Psi(x)=\frac{m_{0}}{\hbar}\Psi(x)
\label{Equation1}
\end{equation}
where the $m_{0}$ and $\Psi\left(x\right)$=$\left(\psi_{+},\psi_{0},\psi_{0},\psi_{-}\right)^{T}$ are the mass and the wave function of the massive vector boson, respectively \cite{16a,16,15a,15b,15bb,15c,15cc,sym,castro,unal}. $\beta^{\mu}(x)$ and $\Sigma_{\mu}(x)$ are Kemmer matrices in 2+1 dimensional curved spacetime and spin
connection coefficients, respectively, and their expressions are given as
\begin{eqnarray}
\beta ^{\mu }(x)=\overline{\mathbf{\sigma }}^{\mu }(x)\otimes I+I\otimes \overline{\mathbf{\sigma }}^{\mu }(x)\nonumber \\
\Sigma _{\mu }\left( x\right) =\Gamma _{\mu }(x)\otimes I+I\otimes \Gamma _{\mu }(x),  \label{connec1}
\end{eqnarray}
in terms of the Dirac matrices, $\overline{\mathbf{\sigma}}^{\mu}(x)$, and spin connections for Dirac particle, $\Gamma _{\mu }(x)$ \cite{15}.

To investigate the quantum gravity effect on the
tunnelling process of the massive vector boson, Eq.(\ref{Equation1}) has to be rewritten in the framework of the GUP.
The simplest and well-known form of the GUP relation is given as follows \cite{24a,24b,feng,li};
\begin{eqnarray}
\Delta x\Delta p\geq\frac{h}{2}\left[1+\alpha(\Delta p)^{2}\right],\label{GUP1}
\end{eqnarray}
where $\alpha$=$\alpha_{0}/M_{p}^{2}$ with Planck mass, $M_{p}$,
and a dimensionless parameter, $\alpha_{0}$. Moreover, the Eq.(\ref{GUP1}) can be
derived by using the modified Heisenberg algebra relation given as \cite{24a,24b,feng,li}
\begin{eqnarray}
\left[ x_{i},p_{j}\right] =i\hbar \delta_{ij}\left[1+\alpha p_{0i}^{2}\right],\label{GUP2}
\end{eqnarray}
with modified position, $x_{i}$, and modified momentum $p_{j}$ operators given in terms of the standard position, $x_{0i}$, and momentum
operators, $p_{0j}$,
\begin{eqnarray}
x_{i} &=&x_{0i},  \nonumber \\
p_{i} &=&p_{0i}(1+\alpha p_{0}^{2}),  \label{GUP3}
\end{eqnarray}
respectively, and $p_{0}^{2}$=$p_{0j}p_{0}^{j}$ \cite{24f}. On the
other hand, the modified energy relation is given by the following form \cite{24aa,24j,24b,feng}:
\begin{eqnarray}
\widetilde{E}=E\left( 1-\alpha E^{2}\right) =E\left[ 1-\alpha \left(p^{2}+m_{0}^{2}\right) \right],  \label{EN}
\end{eqnarray}
with the energy mass shell condition, $E^{2}=p^{2}+m_{0}^{2}$. In addition, the
square of the momentum operator can be expressed after neglecting the higher
order terms of the $\alpha$ parameter as follows \cite{li};
\begin{eqnarray}
p^{2}=p_{i}p^{i}\simeq -\hbar ^{2}\left[ \partial_{i}\partial ^{i}-2\alpha\left( \partial_{j}\partial ^{j}\right) \left( \partial_{i}\partial^{i}\right) \right]. \label{MO}
\end{eqnarray}
Including the Eq.(\ref{GUP3}), Eq.(\ref{EN}) and Eq.(\ref{MO}), the
modified massive vector boson equation can be written as follows;
\begin{eqnarray}
\left( i\beta ^{i}(x)\partial _{i}-i\beta ^{\mu }(x)\Sigma _{\mu }-\frac{m_{0}}{\hbar }\right) \left( 1+\alpha \hbar ^{2}\partial _{j}\partial
^{j}-\alpha m_{0}^{2}\right) \widetilde{\Psi }+i\beta ^{0}(x)\partial _{0}\widetilde{\Psi }=0, \label{MDE1}
\end{eqnarray}
or in its explicit form:
\begin{eqnarray}
i\beta^{0}\partial_{0}\widetilde{\Psi}+\left[i\beta^{i}\left( 1-\alpha m_{0}^{2}\right)\partial_{i}+i\alpha
\hbar^{2}\beta^{i}\partial_{i}\left(\partial_{j}\partial^{j}\right)\right]\widetilde{\Psi} -\frac{m_{0}}{\hbar}\left( 1+\alpha
\hbar^{2}\partial _{j}\partial ^{j}-\alpha m_{0}^{2}\right)\widetilde{\Psi} \nonumber \\-i\beta^{\mu}\Sigma_{\mu}\left( 1+\alpha
\hbar^{2}\partial _{j}\partial ^{j}-\alpha
m_{0}^{2}\right)\widetilde{\Psi}=0,  \label{MDE2}
\end{eqnarray}
where the $\widetilde{\Psi}$ is the modified wave function of a vector boson.

\section{Tunneling of vector boson particle from New-type black hole}

In the framework of the quantum gravity, several studies
have been carried out with a focus on $2+1$ dimensional theories (i.e. a toy
model) \cite{29,25,26,27}. The New Massive Gravity is just one
of them \cite{30}. It has topologically a graviton with mass \cite{30,33,34}. This theory provides black hole solutions. For example, the New-type black hole, which is asymptotically anti-Sitter, static and circular symmetric, is a solution to field equations of the theory. Also, the New-type black hole is conformally flat, and therefore it corresponds to a solution of the 2+1 dimensional conformal gravity theory \cite{35a,35b}. The spacetime metric of the black hole given as
\begin{equation}
ds^{2}=L^{2}\left[ f\left( r\right) dt^{2}-\frac{1}{f\left(r\right)}dr^{2}-r^{2}d\phi^{2}\right], \label{metric}
\end{equation}
where $f(r)=r^{2}+br+c$ and $b$ and $c$ two constant parameters related to the "gravitational hair" and "mass", respectively \cite{35a,35b,14ab,gonz}. $L$ is the AdS$_{3}$ radius defined as $L^{2}$=$\frac{1}{2m^{2}}$=$\frac{1}{2\Lambda}$
in terms of the graviton mass, $m$, (or the cosmological constant, $\Lambda$). The black hole has an outer and inner horizons located at $r_{+}$=$\frac{1}{2}\left(-b+\sqrt{b^{2}-4c}\right)$ and $r_{-}$=$\frac{1}{2}\left(-b-\sqrt{b^{2}-4c}\right)$, respectively. Also, it has a curvature singularity at $r=0$ when $b\neq0$. This singularity is always surrounded by an event horizon. Also, the signatures of the $b$ and $c$ parameters play important role on the mathematical and physical properties of the black hole \cite{35a,14ab}. The horizons coincide with each other when $b^{2}$=$4c$, i.e. the black hole becomes
extremal. Moreover, when $b$=$0$ and $c<0$, the black hole is reduced to
the static Banados-Teitelboim-Zanelli (BTZ) black hole \cite{35a,14ab}.

In this work, the effect of the GUP on the Hawking temperature of the
black hole was investigated by tunneling of the massive vector boson particle. In this investigation, the Kemmer matrices and spin
connection coefficients from \cite{GY3} were employed. Inserting the following ansatz for the modified wave function \cite{16a},
\begin{eqnarray}
\widetilde{\Psi}(x)=\exp \left( \frac{i}{\hbar }S\left( t,r,\phi \right)
\right)\ \left(\begin{array}{c}A\left( t,r,\phi \right) \\B\left(
t,r,\phi \right) \\B\left(
t,r,\phi \right) \\D\left(
t,r,\phi \right)\\ \end{array}\right),  \label{Equation4}
\end{eqnarray}
in Eq.(\ref{MDE2}) and neglecting the terms with $\hbar$, we obtain the following three decoupled differential equations:
\begin{eqnarray*}
B\left[i\frac{L^{2}}{r}\frac{\partial S}{\partial \phi }+i\alpha \frac{1}{r^{3}}\left( \frac{\partial S}{\partial \phi}\right)^{3}-\alpha \frac{\sqrt{f}}{r^{2}}\left( \frac{%
\partial S}{\partial \phi }\right) ^{2}\left( \frac{\partial S}{\partial r}\right)+\alpha L^{2}m_{0}^{2}f\frac{\partial S}{\partial r}-L^{2}\sqrt{f}\frac{\partial S}{\partial r}\right] \\
+B\left[i\alpha \frac{f}{r}\left( \frac{\partial S}{\partial \phi }\right) \left(
\frac{\partial S}{\partial r}\right) ^{2}-\alpha f\sqrt{f}\left(\frac{\partial S}{\partial r}\right)^{3}-i\alpha \frac{m_{0}^{2}L^{2}}{r}\frac{\partial S}{\partial \phi}\right] \\ +A\left[ i\alpha \frac{Lm_{0}}{2r^{2}}\left(
\frac{\partial S}{\partial \phi }\right) ^{2}+i\alpha \frac{Lm_{0}f}{2}\left(\frac{\partial S}{\partial r}\right)^{2}+iL^{3}\frac{m_{0}\left( 1-\alpha m_{0}^{2}\right) }{2}+i\frac{L^{2}}{\sqrt{f}}\frac{\partial S}{\partial t}\right] =0
\end{eqnarray*}
\begin{eqnarray*}
A \left[-L^{2}\sqrt{f}\frac{\partial S}{\partial r}-i\alpha \frac{f}{r}\left( \frac{\partial S}{\partial \phi }\right) \left( \frac{\partial S}{%
\partial r}\right)^{2}+i\alpha \frac{m_{0}^{2}L^{2}}{r}\frac{\partial S}{\partial \phi}-\alpha f\sqrt{f}\left(\frac{\partial S}{\partial r}\right)^{3}-i\alpha \frac{1}{r^{3}}\left( \frac{\partial S}{\partial \phi }\right)^{3}\right] \\
+A\left[-i\frac{L^{2}}{r}\frac{\partial S}{\partial \phi}-\alpha \frac{\sqrt{f}}{r^{2}}\left(\frac{\partial S}{\partial \phi }\right) ^{2}\left( \frac{\partial S}{\partial r}\right)+\alpha L^{2}\sqrt{f}m_{0}^{2}\frac{\partial S}{\partial r}\right]\\
+D\left[i\frac{L^{2}}{r}\frac{\partial S}{\partial \phi }-i\alpha \frac{m_{0}^{2}L^{2}}{r}\frac{\partial S}{\partial \phi }+i\alpha \frac{f}{r}\left( \frac{\partial S}{\partial \phi }\right) \left(
\frac{\partial S}{\partial r}\right)^{2}-L^{2}\sqrt{f}\frac{\partial S}{\partial r}-\alpha f\sqrt{f}\left(\frac{\partial S}{\partial r}\right)^{3}\right] \\
+D\left[\alpha L^{2}m_{0}^{2}f\frac{\partial S}{\partial r}+i\alpha \frac{1}{r^{3}}\left( \frac{\partial S}{\partial \phi }\right) ^{3}-\alpha \frac{\sqrt{f}}{r^{2}}\left( \frac{\partial S}{\partial \phi }\right) ^{2}\left( \frac{\partial S}{\partial r}\right)\right]\\
+B\left[ iL^{3}m_{0}\left( 1-\alpha m_{0}\right) +i\alpha Lm_{0}f\left(\frac{\partial S}{\partial r}\right)^{2}+i\alpha \frac{m_{0}L}{r^{2}}\left(\frac{\partial S}{\partial \phi }\right) ^{2}\right]=0
\end{eqnarray*}
\begin{eqnarray}
B\left[-L^{2}\sqrt{f}\frac{\partial S}{\partial r}-i\frac{L^{2}}{r}\frac{\partial S}{\partial \phi }-\alpha \frac{\sqrt{f}}{r^{2}}\left( \frac{\partial S}{\partial \phi }\right) ^{2}\left( \frac{\partial S}{\partial r}\right)+\alpha L^{2}m_{0}^{2}\frac{\partial S}{\partial r}-i\alpha \frac{1}{r^{3}}\left( \frac{\partial S}{\partial \phi }\right)^{3}\right]\nonumber \\ +B\left[-i\alpha \frac{f}{r}\left( \frac{\partial S}{\partial \phi }\right) \left( \frac{\partial S}{\partial r}\right) ^{2}+i\alpha \frac{m_{0}^{2}L^{2}}{r}\frac{\partial S}{\partial \phi}-\alpha f\sqrt{f}\left( \frac{\partial S}{\partial r}\right)
^{3}\right]\nonumber \\
+D\left[ iL^{3}\frac{m_{0}\left( 1-\alpha m_{0}^{2}\right) }{2}+i\alpha \frac{Lm_{0}}{2r^{2}}\left(\frac{\partial S}{\partial \phi }\right)^{2}+i\alpha \frac{Lm_{0}f}{2}\left(\frac{\partial S}{\partial r}\right)^{2}-i\frac{L^{2}}{\sqrt{f}}\frac{\partial S}{\partial t}\right]=0  \label{Equation5}
\end{eqnarray}
where $A(t,r,\phi)$, $B(t,r,\phi)$ and $D(t,r,\phi)$ are functions of the
spacetime coordinates and $S(t,r,\phi)$ is the particle trajectory known as classical action. The nontrivial solution was found by the condition put on the determinants of the coefficients. Accordingly, we get the modified Hamilton-Jacobi equation as
\begin{eqnarray}
\frac{1}{f}\left(\frac{\partial S}{\partial t}\right) ^{2}-\frac{1}{r^{2}}
\left(\frac{\partial S}{\partial \phi }\right) ^{2}-f\left( \frac{\partial S}{\partial r}\right)^{2}-\frac{m_{0}^{2}L^{2}}{4}+\alpha \left[\frac{9m_{0}^{2}f}{4}\left(\frac{\partial S}{\partial r}
\right) ^{2}+\frac{1}{L^{2}}\left( \frac{\partial S}{\partial t}\right)
^{2}\left(\frac{\partial S}{\partial r}\right)^{2}\right]\nonumber \\+\alpha \left[-\frac{6f}{L^{2}r^{2}}
\left( \frac{\partial S}{\partial r}\right) ^{2}\left( \frac{\partial S}{\partial \phi}\right) ^{2}+\frac{9m_{0}^{2}}{4r^{2}}\left( \frac{\partial S
}{\partial \phi}\right)^{2}+\frac{3m_{0}^{4}L^{2}}{4}-\frac{3f^{2}}{L^{2}}\left(\frac{\partial S}{\partial r}
\right)^{4}-\frac{3}{L^{2}r^{4}}\left( \frac{\partial S}{\partial \phi}
\right)^{4}\right]\nonumber \\+ \alpha \left[\frac{1}{L^{2}r^{2}f}\left(\frac{\partial S}{\partial t}
\right)^{2}\left(\frac{\partial S}{\partial \phi}\right)^{2}-\frac{m_{0}^{2}}{f}\left( \frac{\partial S}{\partial t}
\right)^{2}\right]=0. \label{Equation6}
\end{eqnarray}
Using the separation of variables method, the action function, $S\left(t,r,\phi \right)$, is decomposed into its components as follows;
\begin{eqnarray}
S\left(t,r,\phi\right)=-Et+j\phi +K(r)+C,  \label{Equation7}
\end{eqnarray}
with $E$ and $j$ are the energy and total angular momentum of the particle,
respectively, and $K(r)$=$K_{0}(r)+\alpha K_{1}(r)$ \cite{24k} and the $C$
is a complex constant. Inserting Eq.(\ref{Equation7}) into Eq.(\ref{Equation6}), we get the radial trajectory function, $K\left(r\right)$, of the particle as follows:
\begin{eqnarray}
K_{\pm }(r)=\pm \int \frac{\sqrt{E^{2}-f(m_{0}^{2}+\frac{j^{2}}{r^{2}})}}{f}\left[ 1+\alpha \chi \right] dr \label{Equation8}
\end{eqnarray}
where $\chi$ is
\begin{eqnarray*}
\chi =\frac{1}{4L^{2}f}\left( \frac{5E^{2}L^{2}m_{0}^{2}f-4E^{4}}{E^{2}-f\left( m_{0}^{2}+\frac{j^{2}}{r^{2}}\right) }\right).
\end{eqnarray*}
Here, $K_{+}\left( r\right) $ and $K_{-}\left( r\right)$ represent the
radial trajectories of the outgoing and incoming particles on the outer
horizon, respectively. Therefore, the integration in Eq.(\ref{Equation8}) is calculated as
\begin{eqnarray}
K_{\pm }(r)=\pm i\frac{\pi E}{\left( r_{+}-r_{-}\right) }\left[ 1+\alpha \Omega \right]. \label{Equation9}
\end{eqnarray}
where $\Omega$ is
\begin{eqnarray*}
\Omega=\frac{\left( r_{+}-r_{-}\right) ^{2}\left(
9L^{2}m_{0}^{2}r_{+}^{2}-4j^{2}\right) +16E^{2}r_{+}^{2}}{8L^{2}r_{+}^{2}\left( r_{+}-r_{-}\right)^{2}}.
\end{eqnarray*}
Accordingly, the outgoing and ingoing probabilities of the massive vector boson particles crossing the
outer horizon are given by
\begin{eqnarray}
P_{out}=\exp \left[ -\frac{2}{\hbar } ImK_{+}\left( r\right)
\right]
\nonumber \\
P_{in}=\exp \left[ -\frac{2}{\hbar }ImK_{-}\left( r\right)
\right], \label{Equation10}
\end{eqnarray}
respectively. Furthermore, the total tunneling probability of the massive vector boson particle crossing the horizon is
expressed as \cite{14}
\begin{eqnarray}
\Gamma =e^{-\frac{2}{\hbar }ImS\left( t,r,\phi \right)}=\frac{P_{out}}{P_{in}}.  \label{Equation11}
\end{eqnarray}
Using Eqs.(\ref{Equation10}) and (\ref{Equation11}), and the fact that $ImK_{+}\left( r\right)
=-ImK_{-}\left( r\right) $, the total tunneling probability is obtained as follows;
\begin{eqnarray}
\Gamma=\exp \left[-\frac{4}{\hbar}ImK_{+}\left(r\right)\right]=\exp\left\{-\frac{4\pi E}{\left(r_{+}-r_{-}\right)}\left[ 1+\alpha\Omega\right]\right\}.  \label{Equation12}
\end{eqnarray}
On the other hand, the tunneling probability can be expressed in terms of Boltzmann factor as
\begin{equation}
\Gamma =e^{-\frac{2}{\hbar }ImS}=e^{-\beta E}, \label{Equation13}
\end{equation}
where $\beta $ is the inverse of temperature. Finally, the modified Hawking temperature becomes
\begin{eqnarray}
T_{H}^{^{\prime }}=\hbar \frac{\left( r_{+}-r_{-}\right) }{4\pi }\left[
1+\alpha \Omega \right] ^{-1}. \label{MT0}
\end{eqnarray}
If we expand the $T_{H}^{^{\prime }}$ in terms of the $\alpha $ powers and
neglect the higher order terms, the modified Hawking
temperature of the New-type black hole becomes as follows;
\begin{eqnarray}
T_{H}^{^{\prime }}=T_{H}\left[ 1-\alpha \Omega \right], \label{MT1}
\end{eqnarray}
where the $T_{H}=\hbar \frac{\left( r_{+}-r_{-}\right)}{4\pi}$ is the
standard Hawking temperature of the black hole. From the $T_{H}^{^{\prime}}$
expression, we see that the modified Hawking temperature can be related not
only the mass of the black hole, but also to the AdS$_{3}$ radius,
$L$, (and, hence, to the graviton mass) and the properties of the tunnelled
massive vector boson, such as angular momentum, energy and mass. In addition, this result indicates that the modified Hawking temperature caused by tunneling vector boson particle is completely different from that of both scalar and Dirac particles \cite{GY5}. Therefore, in the framework of the GUP, one can say that massive spin-0 scalar, massive spin-1/2 Dirac and massive spin-1 vector boson particles probe the black hole in different manners. On the
other hand, in the case of $\alpha=0$, the modified Hawking temperature is
reduced to the standard temperature obtained by quantum tunneling process
of the point particles with spin-0, spin-1/2 and spin-1, respectively \cite{GY1,GY3}.

The local stability of a black hole can be analyzed by its heat capacity \cite{Cai}. The positive heat capacity means that black hole is locally stable otherwise it is unstable \cite{Dehghani,Hendi,Miao}. The modified heat capacity of a black hole can be calculated from the relation
\begin{eqnarray}
C^{\prime}=\frac{\partial M}{\partial T_{H}^{\prime}} \label{capa0}
\end{eqnarray}
where $T_{H}^{\prime}$ is the modified Hawking temperature given in Eq.(\ref{MT0}), and $M$ is mass of the black hole given as \cite{14ab},
\begin{eqnarray}
M=\frac{(r_{+}^{2}-r_{-}^{2})}{16G} \label{mass}
\end{eqnarray}
with gravitational constant $G$. Hence, using Eqs.(\ref{MT0}) and (\ref{mass}), the modified heat capacity of the black hole calculated as
\begin{eqnarray}
C^{\prime}=\frac{\pi (r_{+}-r_{-})}{2\hbar G}+\alpha\pi\frac{\left[9L^{2}m_{0}^{2}r_{+}^{5}+9L^{2}m_{0}^{2}r_{+}^{3}r_{-}^{2}+4j^{2}r_{+}^{3}+20r_{+}j^{2}r_{-}^{2}\right]}{16\hbar GL^{2}r_{+}^{3}(r_{+}-r_{-})}\nonumber\\ -\alpha\pi\frac{\left[18L^{2}m_{0}^{2}r_{+}^{4}r_{-}+16E^{2}r_{+}^{3}+16j^{2}r_{+}^{2}r_{-}+8j^{2}r_{+}^{3}\right]}{16\hbar GL^{2}r_{+}^{3}(r_{+}-r_{-})}.\label{capa1}
\end{eqnarray}
or
\begin{eqnarray}
C^{\prime}=\frac{\pi (\mathcal{A}-\mathcal{B})}{16\hbar GL^{2}r_{+}^{3}(r_{+}-r_{-})},\label{capa2}
\end{eqnarray}
where the abbreviations $\mathcal{A}$ and $\mathcal{B}$ are
\begin{eqnarray*}
\mathcal{A}&=&8L^{2}r_{+}^{5}+8L^{2}r_{+}^{3}r_{-}^{2}+9\alpha L^{2}m_{0}^{2}r_{+}^{5}+4\alpha j^{2}r_{+}^{3}+9\alpha L^{2}m_{0}^{2}r_{+}^{3}r_{-}^{2}+20\alpha j^{2}r_{+}r_{-}^{2},\nonumber \\
\mathcal{B}&=&16L^{2}r_{+}^{4}r_{-}+18\alpha L^{2}m_{0}^{2}r_{+}^{4}r_{-}+16\alpha j^{2}r_{-}r_{+}^{2}+16\alpha E^{2}r_{+}^{3}+8\alpha j^{2}r_{-}^{3},
\end{eqnarray*}
respectively. The points where the heat capacity  vanishes or diverges represent the phase transition points where the black hole undergoes from an unstable state to a stable state. The points, where the heat capacity is vanished, correspond to a first-order phase transition while the diverged points to a second-order phase transition. According to the Eq.(\ref{capa2}), the modified heat capacity diverges at point $r_{+}=r_{-}$. This indicates that the black hole is stabilized by passing through a second-order phase transition. The modified heat capacity is positive for $r_{+}>r_{-}$. Hence in this region, the black hole is said to be locally stable. However, for $r_{+}<\left\vert r_{-}\right\vert$, the black hole locally unstable. On the other hand, for $r_{+}>r_{-}$, the modified heat capacity vanishes at point $\mathcal{A}=\mathcal{B}$. This case indicates that the black hole undergoes a first-order phase transition to become stable. In the absence of the GUP effect the heat capacity becomes as
\begin{eqnarray}
C=\frac{\pi (r_{+}-r_{-})}{2\hbar G}. \label{capa3}
\end{eqnarray}
According to this, it can be said that the heat capacity vanishes at point $r_{+}$=$r_{-}$ whereas there is no point where it diverges. This indicates that the black hole has become stable through only the first-order phase transition.

\section{Summary and Conclusion} \label{conclusions}

In this study, we investigate the quantum gravity effect on the tunneling
massive vector boson from the New-type black hole in the context of the 2+1
dimensional New Massive Gravity. For this, using the GUP
relations, we first modified the massive vector boson equation. Then, using the
Hamilton-Jacobi approach, the tunneling probability of the massive vector
particle was derived, and subsequently, the modified Hawking temperature of
the black hole was calculated. We also found that the modified Hawking
temperature not only depends on the black hole's properties, but also
depends on the emitted spin-1 vector boson's mass, energy, total angular
momentum. It is also worth to mention that the modified Hawking temperature is seen to be
depended on the mass of graviton in this context. As can be seen from Eq.(\ref{MT1}), the Hawking temperature increases by the total angular momentum of
the tunneled particle while it decreases by the energy and mass of the
tunnelled particle and the graviton mass.

In addition, according to Eq.(\ref{MT1}), we can summarize some important results as follows:

\begin{itemize}
\item If $9r_{+}^{2}(r_{+}-r_{-})^{2}\frac{m_{0}^{2}}{m^{2}}+16E^{2}r_{+}^{2}
>4j^{2}(r_{+}-r_{-})^{2}$, the modified Hawking temperature of
the tunneling vector boson is lower than the standard temperature. However,
when $9r_{+}^{2}(r_{+}-r_{-})^{2}\frac{m_{0}^{2}}{m^{2}}+16E^{2}r_{+}^{2}<4j^{2}(r_{+}-r_{-})^{2}$, the corrected temperature is higher
than the standard temperature. If $9r_{+}^{2}(r_{+}-r_{-})^{2}\frac{m_{0}^{2}}{m^{2}}+16E^{2}r_{+}^{2}=4j^{2}(r_{+}-r_{-})^{2}$, then the contribution
of the GUP effect is canceled, and the modified temperature of the tunneling
vector boson reduce to the standard temperature.
\item As described previously, the New type black hole is reduced to the
static BTZ black hole in the case of $b=0$ and $c<0$. Hence, the modified
Hawking temperature of the static BTZ black hole under the GUP
effect is
\begin{eqnarray*}
T_{H}^{^{\prime }}=T_{H}\left[ 1-\alpha m^{2}\frac{\left[ \frac{9m_{0}^{2}}{2m^{2}}\left\vert c\right\vert -4j^{2}\right] +4E^{2}}{4\left\vert
c\right\vert }\right] ,
\end{eqnarray*}
where $r_{+}=-r_{-}=\sqrt{\left\vert c\right\vert }$ is used and $T_{H}=\hbar \frac{\sqrt{\left\vert c\right\vert}}{2\pi}$ is the standard
Hawking temperature of the static BTZ black hole in the context of the 2+1
dimensional New Massive Gravity theory \cite{GY1}. In this case, the modified
Hawking temperature is higher than the standard Hawking temperature when $4E^{2}+\frac{9m_{0}^{2}}{2m^{2}}\left\vert c\right\vert <4j^{2}$. On the
other hand, as $4E^{2}+\frac{9m_{0}^{2}}{2m^{2}}\left\vert c\right\vert >4j^{2}$, the modified Hawking temperature is lower than the standard Hawking
temperature.
\item In the absence of the quantum gravity effect, i.e. $\alpha =0$, the
modified Hawking temperature is reduced to the standard temperature obtained
by quantum tunneling of the massive spin-0, spin-1/2 and spin-1 point
particles \cite{GY1,GY3}.
\end{itemize}

In this study we also consider the local stability of the black hole under the GUP effect. At that case, we observe that in the absence of quantum gravity effect, the black hole undergoes first-order phase transition only, but, in the presence of the quantum gravity effect, it undergoes to the first and second-order phase transitions to become stable.

\section*{Conflict of Interests}
The authors declare that there is no conflict of interests regarding the publication of this paper.

\section*{Acknowledgements}

This work was supported by Akdeniz University Scientific Research Projects Unit, and the Scientific and Technological Research Council of Turkey (TUBITAK) 1002-QSP (Project No: 116F329).

\end{document}